\documentclass[twocolumn,secnumarabic,amssymb, amsmath, nofootinbib,tightenlines,
nobibnotes, aps, prl,epsfig]{revtex4}
\usepackage{graphicx}
\usepackage{dcolumn}
\usepackage{bm}
\begin{document}
\preprint{APS/123-QED}
\title{Top structure function at the LHeC }

\author{G.R.Boroun}%
 \email{grboroun@gmail.com; boroun@razi.ac.ir }
\altaffiliation{brezaei@razi.ac.ir}
\affiliation{ Physics Department, Razi University, Kermanshah
67149, Iran}
\date{\today}
\begin{abstract}
The proposed linear and nonlinear behavior for the top structure
function at the LHeC is considered. We present the conditions
necessary to prediction the top structure function
$F_{2}^{t}(x,Q^{2})$ with respect to the different predictions for the bahavior of the gluon at low $x$.\\
\end{abstract}
\maketitle

Heavy quark contributions in leptoproduction are an important
subject in quantum chromodynamic (QCD) phenomenology at lower
values of Bjorken $x$. In leptoproduction, the primary graph is
the Photon-Gluon-Fusion (PGF) model where the incident virtual
photon interacts with a gluon from the target nucleon (Fig.1).
Theoretical calculations for producing
$q\overline{q}$($c\overline{c}$ and $b\overline{b}$) are available
for leading order (LO) and next-leading-order (NLO) processes at
HERA (Experiments H1 and ZEUS). In deep inelastic scattering
(DIS), the kinematical region for the photo production is
available  at $Q^{2}{\geq}5 GeV^{2}$. The available data by the
experiments H1 and ZEUS cover a range of photon virtuality of
$5{\leq}Q^{2}{\leq}2000~GeV^{2}$ and Bjorken scaling variable
$0.0002{\leq}x{\leq}0.05$ [1]. In this framework, the processes
$e+p{\rightarrow}e+c\overline{c}(or ~~b\overline{b})+X$ are
sensitive to the gluon density in the proton and allows to test
its universality. The minimum momentum fraction $x_{g}$ of a gluon
within a nucleon to produce a $q\overline{q}$ heavy quark pair is
$x_{g}>\frac{M^{2}_{q\overline{q}}}{2M_{p}k}$, where $k$ is the
photon energy in the nucleon rest frame. Thus the gluon momentum
fraction $x_{g}^{q\overline{q}}$ in photoproduction (Fig.1) has
this behavior for heavy production samples as
$x_{g}^{t\overline{t}}>x_{g}^{b\overline{b}}>x_{g}^{c\overline{c}}$
along with sensitivities to choice of heavy quarks mass.\\
In recent years, both the H1 and ZEUS Collaborations have measured
the charm and beauty components $F_{2}^{c}$ and $F_{2}^{b}$ of the
structure functions at small $x$ [1] that those are directly
related to the growth of the gluon distribution at that region.
Also authors in Refs. [2-8] show connection between the gluon
distribution  and heavy (charm and beauty) structure functions at
small $x$. This search for charm and beauty production at HERA
emphasis  the importance of having a consistent theoretical
framework for heavy flavor production in DIS. Within the
variable-flavor- number scheme (VFNS)[9-10] the charm and beauty
densities arise via the
$g{\rightarrow}c\overline{c}(b\overline{b})$ evolution.\\
In view of the success of the  BGF within the VFNS perturbatively,
we think that it would be interesting to confront with the top
component $F_{2}^{t}$ of the structure function at small $x$ in
the LHeC project [11-12]. A comparison  LHeC experiment with HERA
experiments (H1 and ZEUS) and fixed target experiments (NMC,
BCDMS, E665 and SLAC) is given by in Fig.2 and a brief overview
can be find in Ref.11. The LHeC project is an investigation of the
possibility of colliding an electron beam from a new accelerator
with the existing LHC proton at an ep center of mass energy beyond
$1~TeV$. The LHeC represents an increase in the kinematic reach of
DIS and an increase in the luminosity. However, the BGF model for
$t\overline{t}$ production will remain a phenomenological model to
be a useful tool for studies QCD at a high gluon density where
saturation and other non-linear physics enter in an essential
way.\\
To study the process $ep{\rightarrow}e(t\overline{t})X$  of the
LHeC, that is sensitive to the gluon density in the proton, the
key question is about the radiative corrections to the heavy
flavor production cross sections that are large. For this reason,
the LHeC study with $50~ GeV$ electrons on $7~ TeV$ protons with
$50~ fb^{-1}$ luminosity is simulated [11]. The kinematic region
accessed at LHeC is $0.000002 < x < 0.8$ and $2 < Q^{2} <
100,000~GeV^{2}$.\\
To estimate the top contribution to the structure function we
firstly consider  the gluon momentum fraction carried by charm and
bottom quarks  and then prediction the the top component
$F_{2}^{t}$ of the structure function. The heavy flavor
contributions to the proton/nucleus structure functions at small
$x$ are given by
\begin{eqnarray}
F_{2}^{q}(x,Q^{2},m^{2}_{q})&=&e_{q}^{2}\frac{\alpha_{s}(\mu^{2}_{q})}{2\pi}\int_{1-\frac{1}{a}}^{1-x}dzC_{g,2}^{q}
(1-z,\zeta)\nonumber\\
&& {\times}G(\frac{x}{1-z},\mu^{2}_{q}),
\end{eqnarray}
where $C_{g,2}^{q}$ and $G=xg$ are the coefficients functions
[2-7,13] and the gluon distribution function. In fact, we assume
that the heavy flavor quarks in the proton structure function only
arise from gluon splitting $g{\rightarrow}q\overline{q}$. Here $a
= 1+\xi$ where $\xi{\equiv}\frac{m^{2}_{q}}{Q^{2}}$ and  the
default common value for the factorization and renormalization
scales is $<\mu^{2}_{q}>=4m_{q}^{2}+Q^{2}/2$. The expansion of the
gluon distribution at an arbitrary point $z=\alpha$ give us an
important equation for obtain the momentum gluon fraction carried
by gluon distributions for create one pair heavy flavor as follows
\begin{eqnarray}
G(\frac{x}{1-\alpha}(K-\alpha),\mu^{2}_{q})={F_{2}^{q}(x,Q^{2},m^{2}_{q})}/{e_{q}^{2}\frac{\alpha_{s}(\mu^{2}_{q})}{2\pi}I},
\end{eqnarray}
where $K=1+\frac{J}{I}$, $ I=\int_{1-\frac{1}{a}}^{1-x}C_{g,2}^{q}
(1-z,\zeta)dz$, $
J=\int_{1-\frac{1}{a}}^{1-x}(z-\alpha)C_{g,2}^{q}(1-z,\zeta)dz$
and $\alpha$ has an arbitrary value $0{\leq}\alpha<1$ [14-15].
This equation is a direct relation between the gluon distribution
at $\beta x$ (where $\beta=\frac{K-\alpha}{1-\alpha}$) and heavy
flavor structure functions ($F_{2}^{c}, F_{2}^{b}$ and
$F_{2}^{t}$) at $x$ value at LO up to NLO analysis. To test the
validity and  obtained the momentum fraction of the gluon
distribution for the top structure function, let us first discuss
this ratio for the charm and beauty structure functions. In Table
1 we obtained the parameter $\alpha$, where describing the
fraction of the gluon momentum in BGF processes for heavy flavor
production. In this approximation, we use the charm and beauty
structure functions obtained by H1 collaboration 2010 [1] and the
gluon distribution is usually taken from the GRV, CETQ or MRST
parametrizations [13]. In Figs.3 and 4 we observe the behavior of
the charm and beauty structure functions with respect to $\alpha$
in average of $x$ values from the H1 data. For any values of these
fractional of the momentum, the charm and beauty structure
functions have values between $0<F_{2}^{c}<0.27$ and
$0<F_{2}^{b}<0.022$ with respect to the experimental data
respectively. According to the Fig.5, we can find the average of
theses momentum fractional as we have $<\alpha^{c}>{\approx}0.1$
and $<\alpha^{b}>{\approx}0.7$.\\
Let us determine the predictions for $<\alpha^{t}>$. We know that
the top quark can be produced the LHeC from $Wb {\rightarrow}~t$
reactions where the b quark arises from the intrinsic beauty
component. But we are interesting to  produce top-quark pairs in
$\gamma^{*}p{\rightarrow}t\overline{t}X$ reactions. The average
value $\alpha^{t}$ have to go to beyond 1 value as can be seen
from the ratio of intrinsic beauty to intrinsic top scales  as
$\frac{m^{2}_{b}}{m^{2}_{t}}{\simeq}\frac{1}{1500}$ where the
ratio of intrinsic charm to intrinsic beauty scales is
$\frac{m^{2}_{c}}{m^{2}_{b}}{\simeq}\frac{1}{10}$. But the
asymptotic value  $\alpha$ is 1 for fractional of gluon momentum,
therefore the threshold Bjorken scaling for top pair  production
reaches to low values of $x$ at fixed  $Q^{2}$ (Table 2). In this
table, we observe that as $Q^{2}$ increase, the initial threshold
for the Bjorken scaling  for $0<\alpha<1$ increases. In Fig.6 we
show our predictions for the top structure function $F_{2}^{t}$
according to the Bjorken scaling threshold as a function of
$\alpha$ at $Q^{2}$ and $x$ constant values. We can observe that
top structure functions increase as $Q^{2}$ increase and $x$
decreases. But with respect to the average $\alpha$ for charm and
beauty, we expect that $<\alpha^{t}>{\rightarrow}~1$. Therefore we
choose this expanding point for top pair production to be
${\simeq}0.95$. As Fig.7 shows, the NLO predictions for the top
structure function are rather stable for $Q^{2}$ of order of
$m^{2}_{t}$ at low-$x$ values under the renormalization scale with
respect to the luminosity. At that expanding point, the top
structure functions obtained as a function of $x$ for $Q^{2}=10,
100, 1000$ and $10000~GeV^{2}$. We observe that the initial point
of $x$ increase as $Q^{2}$ increases. This value has a maximum at
$x{\sim}0.2$ when $Q^{2}$ increase to $100000~GeV^{2}$, but this
maximum point can be increase toward $0.8$ when the expanding
point decreases toward $0$. With respect to the luminosity in the
LHeC, we have expect that the available data to be  over a wide
small-x range. Thus one can better explore the small-x region
where non-linear evolution is required as $\ln 1/x$ terms in the
evolution become important [ 16-17] and where resummation
approaches may be required [18]. The main characteristic of this
nonlinear evolution equation is that it predicts a saturation of
the gluon distribution at very small $x$, which the recombination
processes such as $gg {\rightarrow} g$, leading to non-linear
evolution equation. This saturation effects may be applied to the
top structure function  at very small $x$ values, which may be
possible by studying the very small x region at somewhat larger
$Q^{2}$ at the LHeC. Now we predict the saturation effects to the
top structure function $F^{t}_{2}(x,Q^{2})$ in the LHeC kinematic
range. This picture allows us to write the GLRMQ [19] equation for
the gluon structure function at small $x$ as follows
\begin{eqnarray}
\frac{{\partial}G(x,Q^{2})}{{\partial}{\ln}Q^{2}}=\frac{{\partial}G(x,Q^{2})}{{\partial}{\ln}Q^{2}}|_{DGLAP}\nonumber\\
-\frac{{\gamma}\alpha^{2}_{s}(Q^{2})}{R^{2}Q^{2}}\int^{1}_{x}\frac{dz}{z}[G(\frac{x}{z},Q^{2})]^{2}.
\end{eqnarray}
The factor $\gamma$ found to be $=\frac{81}{16}$ for $N_{c}=3$,
and the first term in the r.h.s. is the usual linear DGLAP term in
DLLA and the second term is nonlinear in gluon density. Here $R$
is the correlation radius between two interacting gluons and $\pi
R^{2}$ is the target area where gluons inhabit. The value of $R$
depends on how the gluon ladders couple to the proton, or on how
the gluons are distributed within the proton. $R$ will be of the
order of the proton radius $(R\simeq5\hspace{0.1cm} GeV^{-1})$ if
the gluons are spread throughout the entire nucleon, or much
smaller $(R\simeq2\hspace{0.1cm} GeV^{-1})$ if gluons are
concentrated in hot- spot [20] within the proton. This nonlinear
evolution equation can be solve for the nonlinear gluon
distribution behavior by some methods as those are presented in
Refs.[21,22]. Recently, a general solution of the gluon density
was performed in the nonlinear evaluation equation kinematical
region [23], as the gluon distribution in terms of the initial
condition can be expressed by
\begin{eqnarray}
G(x,Q^{2})&=&[G(x,Q^{2})|_{Linear-DGLAP}]-\int_{\chi}^{1}G^{2}(z,Q_{0}^{2})Fe^{Y}\nonumber\\
&&{\times}BesselI(0,2\sqrt{U}\sqrt{{\ln}\frac{z}{x}})\frac{dz}{z},
\end{eqnarray}
where
\begin{eqnarray}
G(x,Q^{2})|_{Linear-DGLAP}=e^{\eta(Q^{2})}[G(x,Q_{0}^{2})+\int_{x}^{1}G(z,Q_{0}^{2}){\times}\nonumber\\
\frac{\sqrt{\zeta}}{\sqrt{{\ln}\frac{z}{x}}}BesselI(1,2\sqrt{\zeta}\sqrt{{\ln}\frac{z}{x}})\frac{dz}{z}].
\end{eqnarray}
Here
$\eta(Q^{2})=\int_{Q_{0}^{2}}^{Q^{2}}\frac{1}{{\ln}\frac{Q^{2}}{\Lambda^{2}}}d
{\ln}Q^{2}$and $ {\zeta} {\equiv}
\frac{12\eta(Q^{2})}{\beta_{0}}$, other parameterizes can be find
in Ref.[23]. Also $\chi=\frac{x}{x_{0}}$, where $x_{0}(=0.01)$ is
the boundary condition that the gluon distribution joints smoothly
onto the
unshadowed region.\\
Fig.8 represent our prediction results  for the top structure
function nonlinear behavior for $R=5~GeV^{-1}$ at
$Q^{2}=10000~GeV^{2}$. This result shows that the top structure
function behavior is tamed with respect to nonlinear terms at the
GLR-MQ equation. The differences are not large for $1E-5<x<1E-2$
but these differences are more concretely for $1E-7<x<1E-5$. It
shows that screening effects are provided by a multiple gluon
interaction, which leads to the nonlinear terms in the DGLAP
equation. For $x<1E-7$, the nonlinear behavior shows that the top
structure function fall deeply to ward negative values. May be the
Pomeron amplitude have to fixed to the top structure function at
this region.\\
In conclusion, we prediction  the top structure function at the
LHeC project at low $x$ and high $Q^{2}$ values. At low $x$ we
expected that extension of the conventional QCD DGLAP resummation
is necessary to explain the data, but the nonlinear behavior tamed
deeply at very low $x$ as may be add possible scenarios such as
BFKL resummation. We observed that, as $x$ decreases, the
singularity behavior of the top structure function is tamed by
shadowing effects.\\
\subsection{Acknowledgment}
Author thanks  Thomas Gehrmann for discussions which prompted this
study and the Department of Physics of the University of
Zurich for their warm hospitality. We are also grateful to A.Cooper-Sarkar for critically reading the manuscript and important comments.\\

\textbf{References}\\
1. F.D. Aaron et al. [H1 Collaboration],Phys.Lett.b\textbf{665},
139(2008), Eur.Phys.J.C\textbf{65},89(2010); J. Breitweg et. al.,
[ZEUS Collaboration], Eur. Phys. J. C\textbf{12}, 35 (2000); S.
Chekanov et. al., [ZEUS Collaboration], Phys. Rev. D\textbf{69},
012004 (2004); A.Aktas et al. [H1 Collaboration],
Eur.Phys.J.C\textbf{45}, 23 (2006); Eur.Phys.J.C\textbf{40}, 349
(2005).\\
2. N.N.Nikolaev and V.R.Zoller, Phys.Atom.Nucl\textbf{73},
672(2010); Phys.Lett.B \textbf{509}, 283(2001);  N.N.Nikolaev,
J.Speth and V.R.Zoller, Phys.Lett.B\textbf{473}, 157(2000);
R.Fiore,
N.N.Nikolaev and V.R.Zoller, JETP Lett\textbf{90}, 319(2009).\\
3.A.V.Kotikov, A.V.Lipatov, G.Parente and N.P.Zotov, Eur.\
Phys.\ J.\  C {\bf 26}, 51 (2002).\\
4. A.Y.Illarionov, B.A.Kniehl and A.V.Kotikov, Phys.\ Lett.\
B {\bf 663}, 66 (2008); A. Y. Illarionov and A. V. Kotikov, Phys.Atom.Nucl. {\bf75}, 1234 (2012).\\
5. N.Ya.Ivanov, and B.A.Kniehl, Eur.Phys.J.C\textbf{59}, 647(2009);  N.Ya.Ivanov, Nucl.Phys.B\textbf{814}, 142(2009); Eur.Phys.J.C{\bf59}, 647(2009)\\
6. I.P.Ivanov and N.Nikolaev,Phys.Rev.D{\bf65},054004(2002).\\
7. N.N.Nikolaev and V.R.Zoller, Phys.Lett. B\textbf{509},
283(2001); Phys.Atom.Nucl.\textbf{73}, 672(2010); V.R.Zoller,
Phys.Lett. B\textbf{509},
69(2001).\\
8. G.R.Boroun, B.Rezaei, JETP,Vol.115, No.7, PP.427 (2012);
Nucl.Phys.B{\bf857}, 143(2012); Eur.Phys.J.C{\bf72}, 2221 (2012);
EPL{\bf100},41001(2012); Nucl.Phys.A{\bf929}, 119(2014); G.R.Boroun, Nucl.Phys.B{\bf884}, 684(2014).\\
9. M.A.G.Aivazis, et.al., Phys.Rev.D\textbf{50},
3102(1994).\\
10. J.C.Collins, Phys.Rev.D\textbf{58},
094002(1998).\\
11. P.Newman, Nucl.Phys.Proc.Suppl.{\bf191}, 307(2009);
S.J.BRODSKY, hep-ph/arXiv:1106.5820 (2011); Amanda Cooper-Sarkar,
hep-ph/arXiv:1310.0662 (2013).\\
12. LHeC Study group, CERN-OPEN-2012-015; F. D. Aaron et al. [H1
and ZEUS Collaboration], JHEP {\bf1001},
109(2010)[hep-ex/arXiv:0911.0884
]; LHeC Study group, LHeC-Note-2012-005 GEN.\\
13. M.Gluk, E.Reya and A.Vogt, Z.Phys.C\textbf{67}, 433(1995); Eur.Phys.J.C\textbf{5}, 461(1998); H.L.Lai et. al., [CTEQ Collaboration],
 Eur.Phys.J.C\textbf{12}, 375(2000); A.D.Martin, R.G.Roberts, W.J.Stirling and R.S. Thorn,
Eur.Phys.J.C\textbf{35}, 325(2004); A.D. Martin, W.J. Stirling, R.S. Thorne and G. Watt, Eur.Phys.J.C{\bf63}, 189(2009).\\
14. A.M.Cooper-Sarkar et.al., Z.Phys.C\textbf{39},
 281(1998); A.M.Cooper-Sarkar and R.C.E.Devenish, Acta.Phys.Polon.B\textbf{34},
 2911(2003).\\
15. G.R.Boroun and B.Rezaei, Eur.Phys.J.C{\bf72}, 2221(2012); JETP {\bf115}, 427(2012).\\
16. V. Fadin, E. Kuraev and L. Lipatov, Sov. Phys. JETP 44, 443
(1976); V. Fadin, E. Kuraev and L. Lipatov, Sov. Phys. JETP 45,
199 (1977); Y. Balistsky and L. Lipatov, Sov. J. Nucl. Phys. 28,
822 (1978).\\
17. M. Ciafaloni, Nucl. Phys.B {\bf 296}, 49 (1988); S. Catani, F.
Fioriani and M. Marchesini, Phys. Lett.B {\bf 234}, 339 (1990); S.
Catani, F. Fioriani and M. Marchesini, Nucl. Phys.B {\bf 336}, 18
(1990); M. Marchesini, Nucl. Phys.B {\bf 445}, 49 (1995).\\
18. G. Altarelli, R. Ball and S. Forte, Nucl. Phys.B {\bf 799},
199(2008).\\
19. A.H.Mueller and J.Qiu, Nucl.Phys.B\textbf{268}, 427(1986);
L.V.Gribov, E.M.Levin and M.G.Ryskin, Phys.Rep.\textbf{100},
 1(1983).\\
20. E.M.Levin and M.G.Ryskin, Phys.Rep.\textbf{189}, 267(1990).\\
21. G.R.Boroun, Eur.Phys.J.A \textbf{42}, 251(2009).\\
22. M.Devee and J.K.Sarma, Eur.Phys.J.C \textbf{74}, 2751(2014); Nucl.Phys.B \textbf{885}, 571(2014).\\
23. G.R.Boroun and S.Zarrin, Eur.Phys.J.Plus \textbf{128}, 119(2013).\\

\begin{figure}
\includegraphics[width=0.2\textwidth]{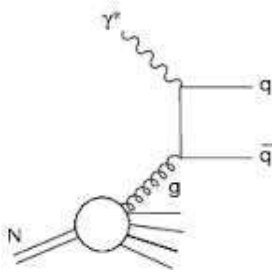}
\caption{ Boson-gluon-fusion (BGF) graph. }\label{Fig1}
\end{figure}
\begin{figure}
\includegraphics[width=0.3\textwidth]{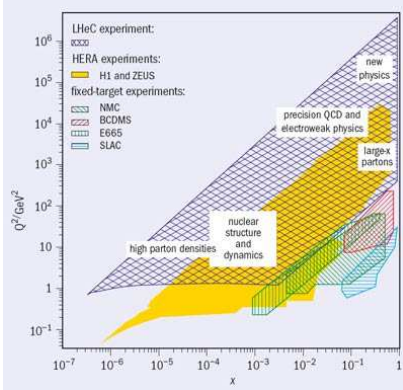}
\caption{ Kinematic plane for ep collisions in Bjorken-$x$ and
resolving power $Q^{2}$ showing the coverage of fixed target
experiments, HERA and an LHeC. }\label{Fig1}
\end{figure}
\begin{figure}
\includegraphics[width=1\textwidth]{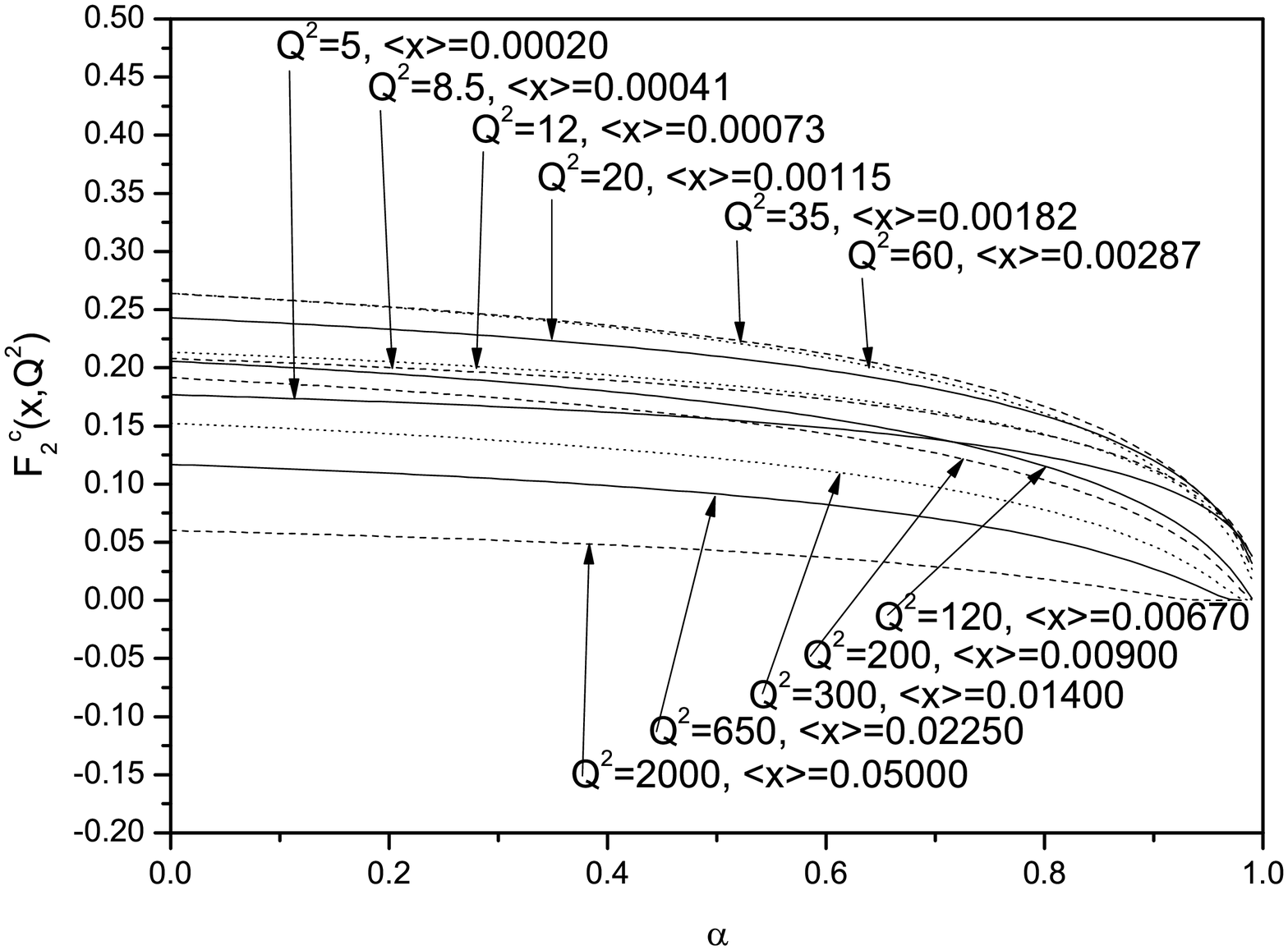}
\caption{ The charm structure function behavior versus  $\alpha$
for the fixed-$Q^{2}$ and averages $<x>$ values according to the
H1 Collab. data (2010)[1]. }\label{Fig1}
\end{figure}
\begin{figure}
\includegraphics[width=1\textwidth]{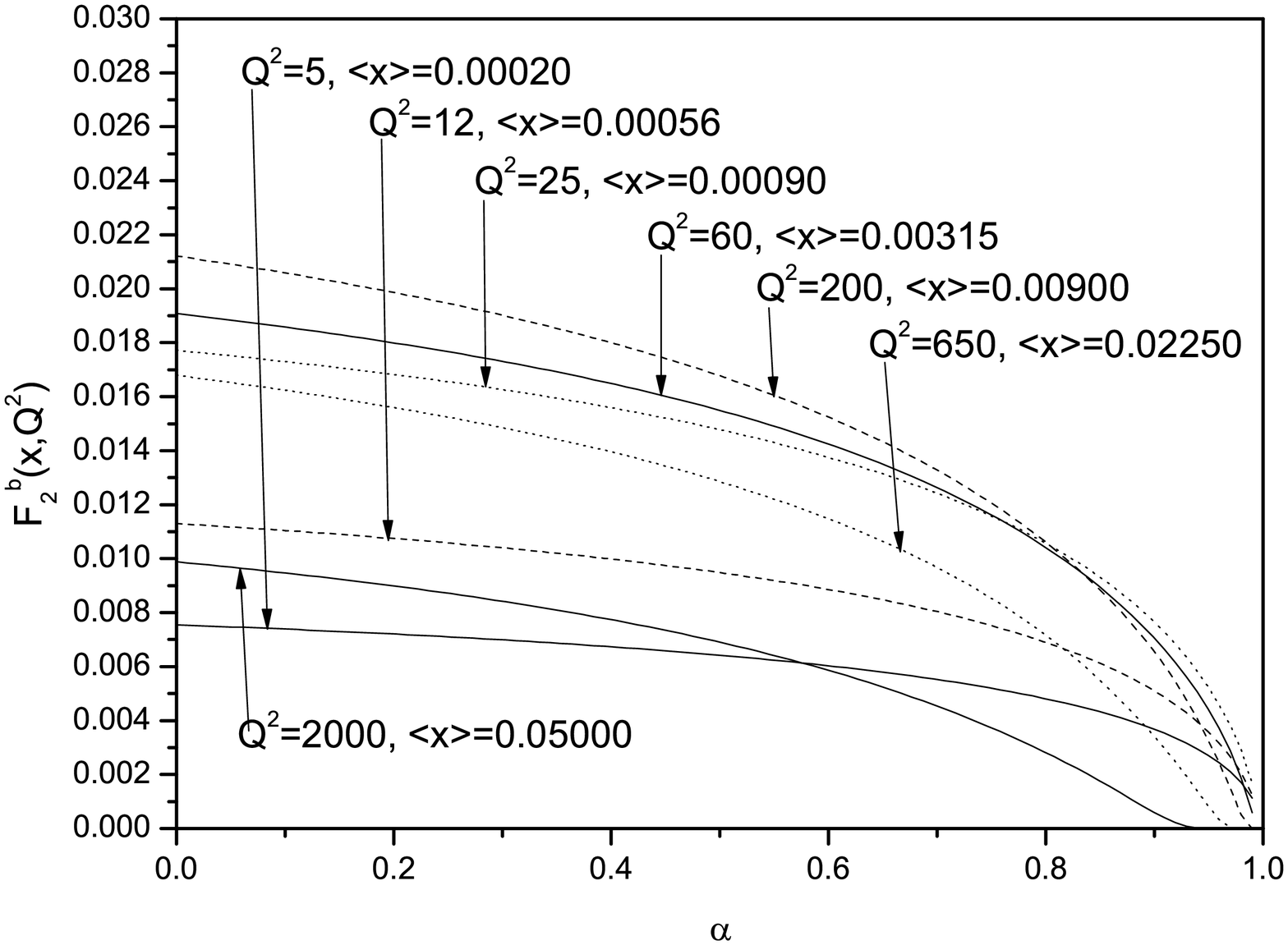}
\caption{ The beauty structure function behavior versus  $\alpha$
for the fixed-$Q^{2}$ and averages $<x>$ values according to the
H1 Collab. data (2010)[1]. }\label{Fig1}
\end{figure}
\begin{figure}
\includegraphics[width=1\textwidth]{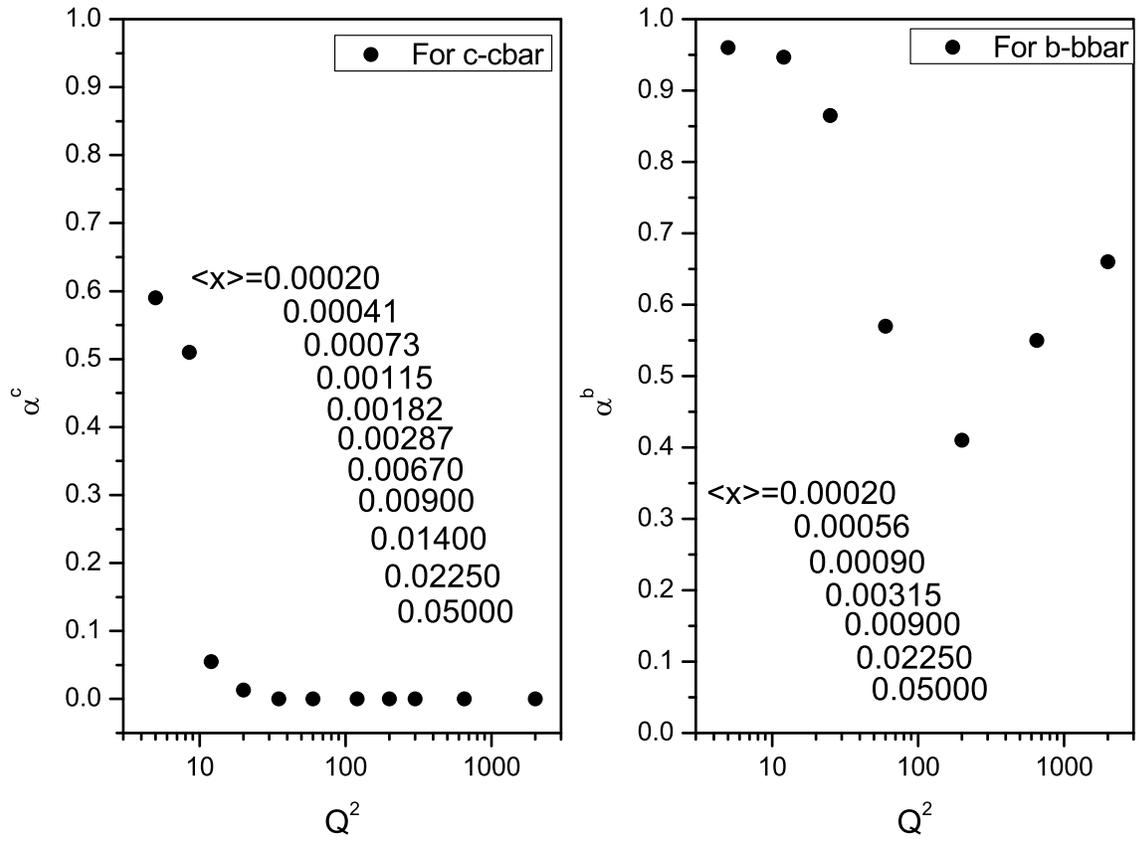}
\caption{ The expanding point $\alpha$ for the charm and beauty
structure functions versus $Q^{2}$ at averages $<x>$ according to
the H1 Collab. data (2010)[1]. }\label{Fig1}
\end{figure}
\begin{figure}
\includegraphics[width=1\textwidth]{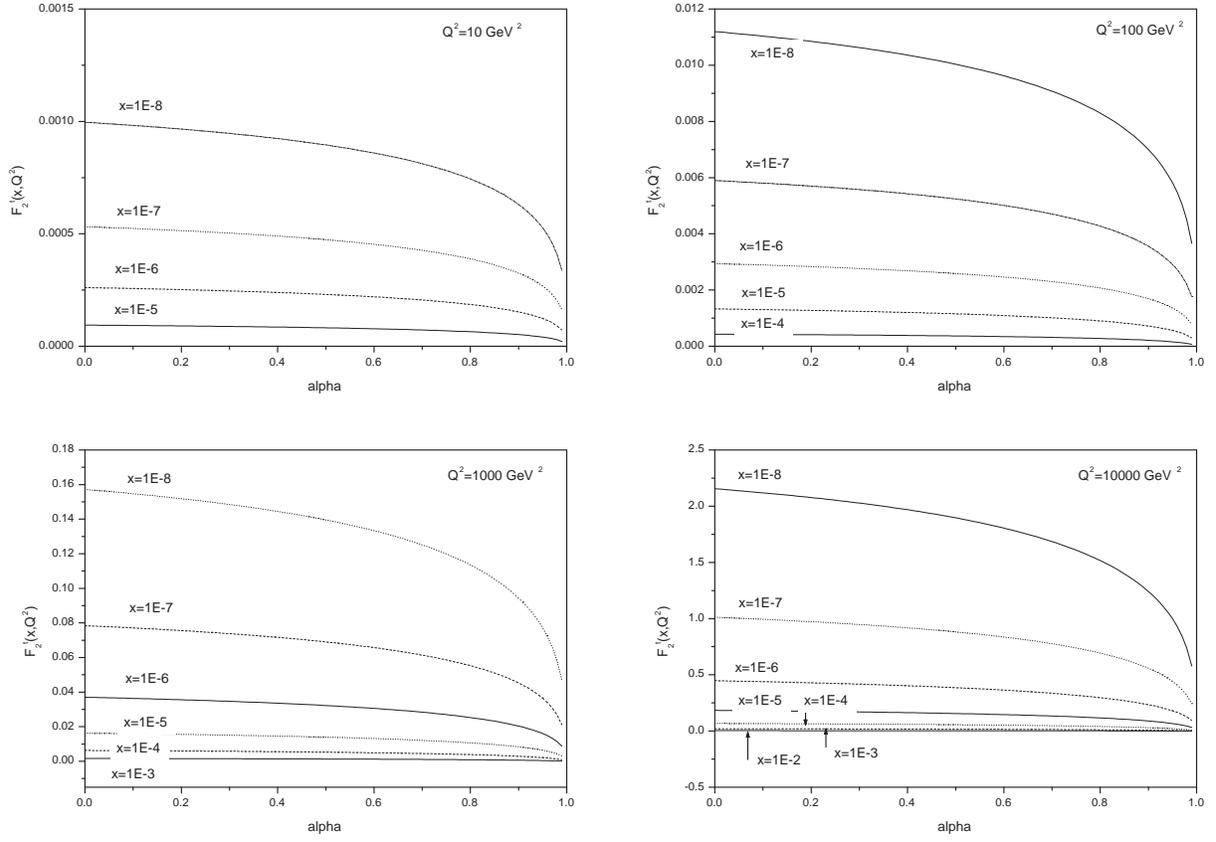}
\caption{ The top structure function behavior predictions versus
$\alpha$ for the fixed-$Q^{2}$. }\label{Fig1}
\end{figure}
\begin{figure}
\includegraphics[width=1\textwidth]{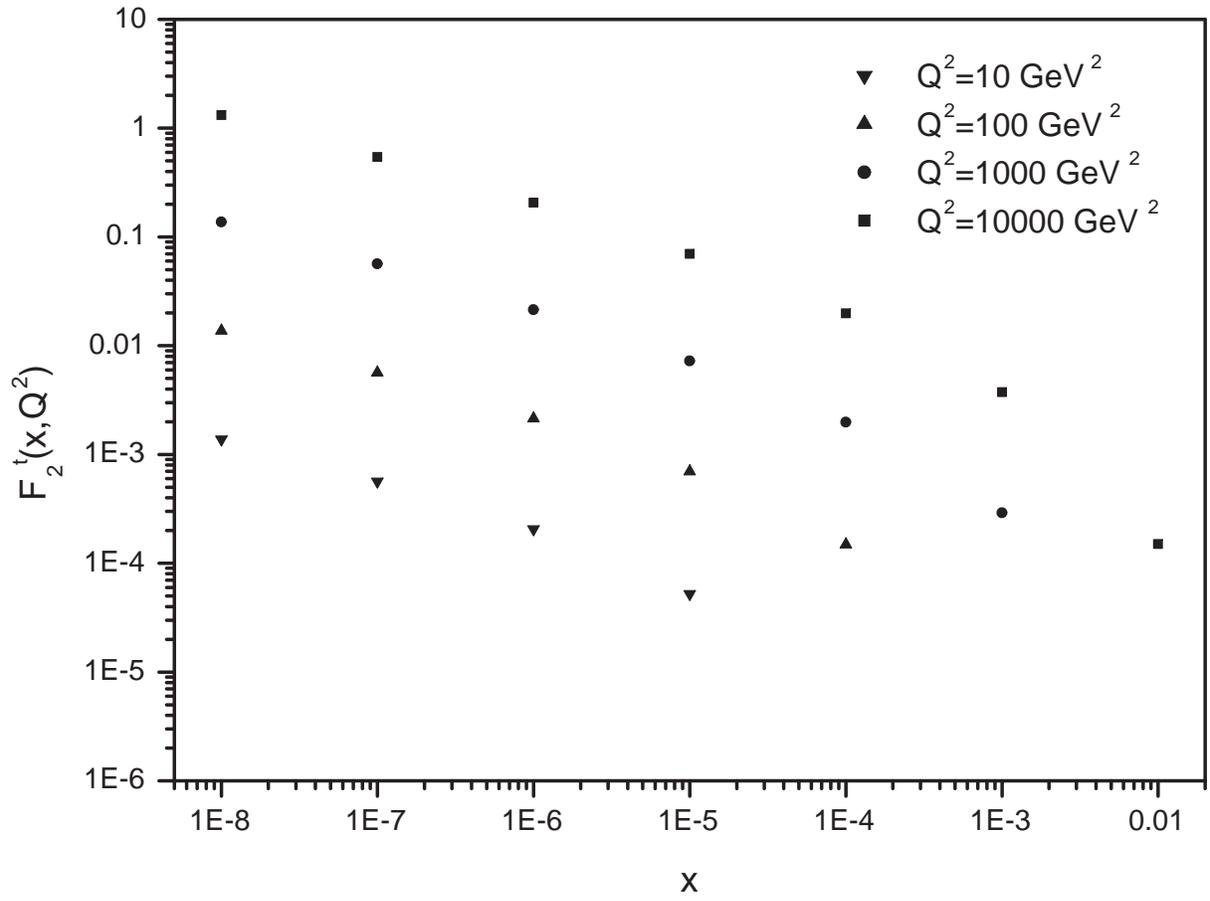}
\caption{ Our predictions for top structure function in the LHeC
for $\alpha=0.95$ at $Q^{2}=10000~GeV^{2}$. }\label{Fig7}
\end{figure}
\begin{figure}
\includegraphics[width=1\textwidth]{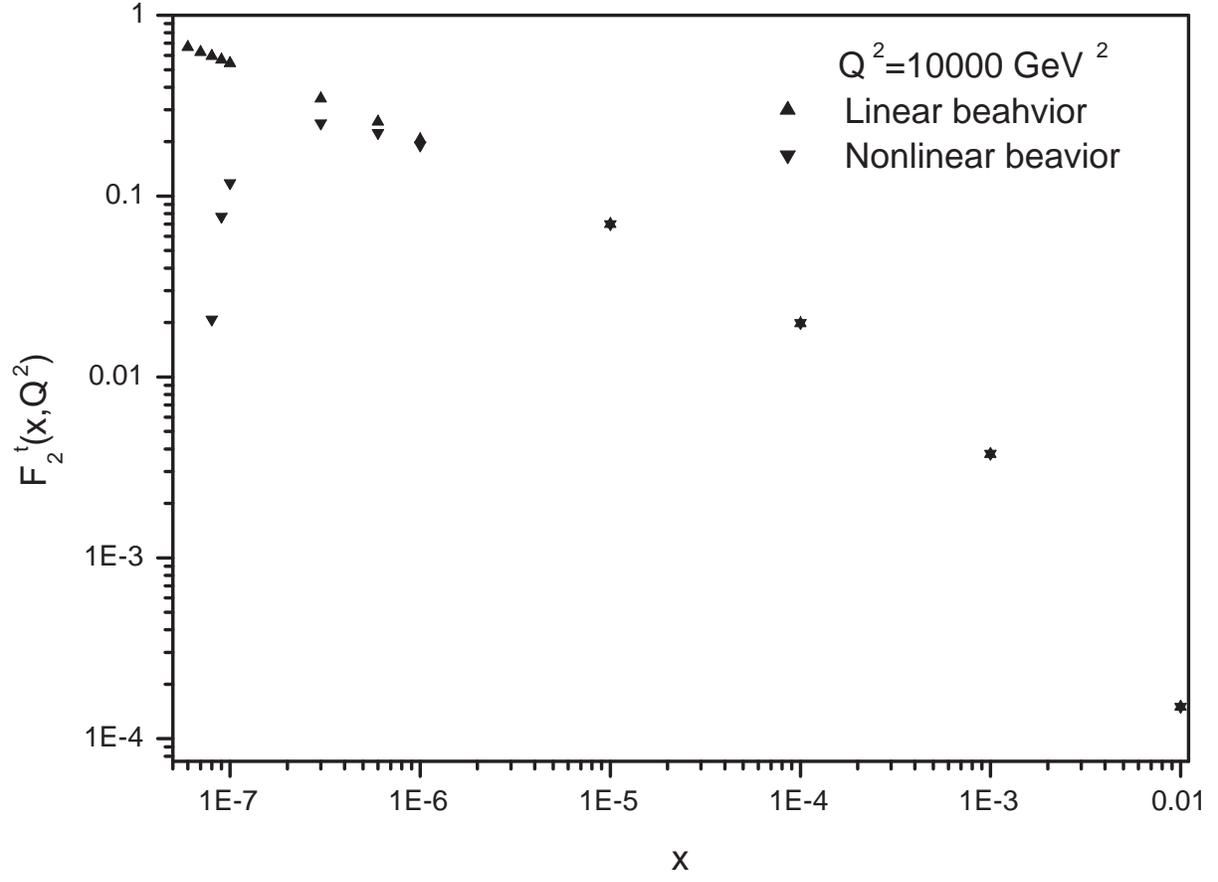}
\caption{ Nonlinear behavior for top structure function at
$Q^{2}=10000~GeV^{2}$. }\label{Fig8}
\end{figure}

\begin{table}
\centering \caption{ Values of the gluon momentum fraction
($\alpha$) with respect to the charm and beauty structure
functions (H1 Collab.2010 [10]) at the average values of $x$
($<x>$) for $<\mu^{2}_{q}>$. }\label{table:table1}
\begin{minipage}{\linewidth}
\renewcommand{\thefootnote}{\thempfootnote}
\centering
\begin{tabular}{|l||c|c|c|c|c|c|l|} \hline\noalign{\smallskip} $Q^{2}(GeV^{2})$ & $ <x>$ &
$ <F_{2}^{c}{\pm}\delta>$ & $ \alpha^{c}$  & $<x>$
&$ <F_{2}^{b}{\pm}\delta>$ & $ \alpha^{b} $  \\
\hline\noalign{\smallskip}
5 & 0.00020 & 0.1490{$\pm$}0.010 & 0.590 & 0.00020 & 0.00244{$\pm$}0.010 &0.960  \\
8.5 &0.00041 & 0.1815{$\pm$}0.010 & 0.510 & - & - &-   \\
12 & 0.00073 & 0.2115{$\pm$}0.010 & 0.055 & 0.00056 & 0.00369{$\pm$}0.011 &0.947    \\
20 & 0.00115 & 0.2424{$\pm$}0.010 & 0.013 & - & - &-   \\
25 & - & - & - &  0.00090 & 0.00896{$\pm$}0.011 &0.865  \\
35 & 0.00182 & 0.2692{$\pm$}0.011 & {$\simeq$}~0 & - & - &- \\
60 & 0.00287 & 0.2772{$\pm$}0.010 & {$\simeq$}~0 & 0.00315 & 0.01467{$\pm$}0.010 &0.570  \\
120 & 0.00670 & 0.2430{$\pm$}0.017 & {$\simeq$}~0 & - & - &-  \\
200 & 0.00900 & 0.2015{$\pm$}0.028 & {$\simeq$}~0 & 0.00900 & 0.01782{$\pm$}0.028 &0.410    \\
300 & 0.01400 & 0.1975{$\pm$}0.029 & {$\simeq$}~0 & - & - &- \\
650 & 0.02250 & 0.1440{$\pm$}0.033 & {$\simeq$}~0 & 0.02250 & 0.01210{$\pm$}0.033 &0.550  \\
2000 & 0.05000 & 0.0600{$\pm$}0.043 & {$\simeq$}~0 & 0.05000 & 0.00511{$\pm$}0.043 &0.660  \\

\hline\noalign{\smallskip}
\end{tabular}
\end{minipage}
\end{table}
\begin{table}
\centering \caption{ The Bjorken scaling initial threshold for
 top pair production at the LHeC. }\label{table:table1}
\begin{minipage}{\linewidth}
\renewcommand{\thefootnote}{\thempfootnote}
\centering
\begin{tabular}{|l|c|} \hline\noalign{\smallskip} $Q^{2}(GeV^{2})$ & $ x_{0}(x{<}x_{0})$  \\
\hline\noalign{\smallskip}
10 & 0.0001 \\
100 & 0.0010 \\
1000 & 0.0100 \\
10000 & 0.1000 \\
\hline\noalign{\smallskip}
\end{tabular}
\end{minipage}
\end{table}
\end{document}